\documentclass[aps,prb, preprint, superscriptaddress,showpacs,showkeys,floatfix]{revtex4-1}
\usepackage{amsmath}
\usepackage{bm}
\usepackage{amssymb}
\usepackage{graphicx}
\usepackage{epstopdf}
\usepackage{epsfig}
\usepackage{color}
\usepackage{hyperref}
\usepackage{float}
\usepackage{braket}
\usepackage{lipsum}

\begin{document}

\title{Coherent light generation in hybrid atomic-nanophotonic integrated structures}

\author{Hadiseh Alaeian}
\affiliation{Department of Physics and Astronomy, Northwestern University, Evanston, IL 60208, USA}
\author{Brian C. Odom}
\affiliation{Department of Physics and Astronomy, Northwestern University, Evanston, IL 60208, USA}
\author{Jorge Bravo-Abad}
\affiliation{Departamento de F\'{i}sica Te\'{o}rica de la Materia Condensada  and Condensed Matter Physics Center (IFIMAC), Universidad Aut\'{o}noma de Madrid, E-28049 Madrid, Spain}

\date{\today}

\begin{abstract}
The integration of neutral atoms with nanophotonic structures offer significant potential as a versatile platform to explore fundamental light-matter interactions as well as realizing novel quantum-optical devices. Here, we investigate the possibility of creating low-threshold micro-scale lasers in hybrid systems based on integrating room-temperature atomic gases with both dielectric and metallic nanophotonic systems. We particularly focus on studying two different devices resulting from incorporating an optically-pumped Rb-ethane mixture in a dielectric ring resonator and a plasmonic lattice. We show in both cases the combination of the optical gain provided by the atomic vapor, along with the unique field-confinement properties of nanophotonic structures, enables generating of coherent radiation, i.e. laser light, at low power levels. In addition, we provide general design guidelines for these hybrid nano-lasers and an efficient density matrix-based formalism for studying these systems. Our results demonstrate a unique route towards small foot-print, highly efficient, and fast lasers, which paves the way towards the development of a whole new class of active nanophotonic and metamaterial systems.
\end{abstract}

\maketitle

\section{Introduction}

The field of quantum optics has benefited a lot from theoretical and technological advances in solid-state photonics. With the first demonstration of single photon sources in a nano-device at the beginning of this century~\cite{Santori02, Michler00, Moreau01, Solomon01, Santori01}, these systems became one of the prominent platforms for realizing various quantum optical phenomena and devices. Low-threshold photonic crystal lasers~\cite{Noda01, Wu04, Altug05}, plasmonic nanolasers~\cite{Bergman2003,Noginov2009,Oulton2009,Hess2012}, strong coupling~\cite{Spillane05, Yoshie04, Hennessy07}, photon blockade~\cite{Bajcsy13, Muller15, Majumdar13}, chiral quantum optics and entanglement~\cite{Lodahl16, Reiserer15, Lang17, Mahmoodian17} are some of intriguing quantum optical phenomena studied and realized in solid-state platforms. In spite of their appealing features for integration as a monolithic module, the large crystal-field broadening and heterogeneity of quantum dots have been two major bottlenecks of solid-state systems in achieving long coherence length and large-scale quantum networks~\cite{Kalb17}.

Meanwhile in parallel, substantial developments in the other disciplines of physics have been made toward realizing different quantum information processing schemes. In particular, In the realm of atomic, molecular, and optical physics (AMO) trapped ions and neutral atoms have been employed as high-fidelity nodes and qubits in scalable quantum computers~\cite{Kaler03, Leibfried03, Benhelm08, Pinkse00, Kuzmich03}. So far, AMO systems appear to be the most ideal known candidates regarding their coherence features, with very well-established and precise quantum state manipulation schemes. On the other hand, thanks to the advances in nano-technology and fabrication techniques, solid-state nanophotonics has become the most suitable platform for integration and scalibility. Therefore, a new promising approach could benefit from the best of two worlds, in a hybrid scheme, on an interface between AMO and solid-state photonics~\cite{Wallquist09}.

Among different candidates in AMO, neutral atoms are easier to work with, since the presence of a device in the vicinity of the ions would substantially perturb the ion traps due to the induced surface charges. Moreover, in spite of some successful demonstration of cold atom-photonic device couplings the miniaturization of the typical cold atom setups compatible with nano-devices is still a challenge~\cite{Du04,Aoki06,Tiecke14}. Unlike cold atoms, thermal vapors lend themselves for proper integration and coupling to the photonic modes as has been demonstrated in hollow-core fibers and waveguides~\cite{Yang07, Epple14, Slepkov10}. Very recently, the thermal vapors of alkali atoms have been successfully combined with on-chip, integrated plasmonic and photonic devices where the atomic transitions have been modified with the photonic modes showing noticeable effect as Fano resonances or sub-Doppler line modifications~\cite{Ritter15,Stern13, Aljunid16}.

In this paper, we investigate the possibility of laser light generation in hybrid systems that combine room-temperature atomic clouds and nanophotonic structures. The optical gain we consider in our study consists of an optically-pumped mixture of Rb and ethane, as has been successfully employed in the past in diode pumped alkali lasers (DPAL) for achieving large optical gain within the infrared~\cite{Konefal99, Krupke03, Krupke12}. This Rb-ethane mixture is incorporated into two different classes of integrated nanophotonic platforms, an all-dielectric ring resonator supporting whispering gallery modes with large quality factor, and a plasmonic lattice, featuring lattice resonances with moderate quality factor and large field enhancement at the subwavelength scale. To tackle this problem, we develop an efficient density matrix-based formalism that properly captures the dynamics and decoherence of light-matter interaction in the studied systems. The method can be efficiently employed to study the transient and steady state features of both the atomic level populations and the electromagnetic field. Using this approach, we show the temporal and spatial field-confinement properties of the considered nanophotonic systems can be tailored to enable laser light emission at low pumping power levels. Furthermore, we derive a simplified model to provide general design guidelines for this novel class of hybrid active systems.

\section{Theoretical framework}
In spite of their very different wavelength, all DPALs share the same working principle. The outer shell of alkalis have three main electronic levels of $n^{2}S_{1/2}, n^{2}P_{1/2},n ^{2}P_{3/2}$. In a mixture with alkali atoms excited at $D_2$-line (i.e. $n^{2}P_{3/2}$), the collision of the excited atoms with buffer gas molecules (typically a hydrocarbon) efficiently transfers part of the population from the excited state to $n^{2}P_{1/2}$. With a properly chosen buffer gas at an optimized pressure a population inversion between $n^{2}P_{1/2}, n^{2}S_{1/2}$ would be built up hence, an optical gain at this transition (known as $D_1$-line) is established~\cite{Krupke12}. 

\begin{figure*}[t]
\centering
\includegraphics[width=\linewidth]{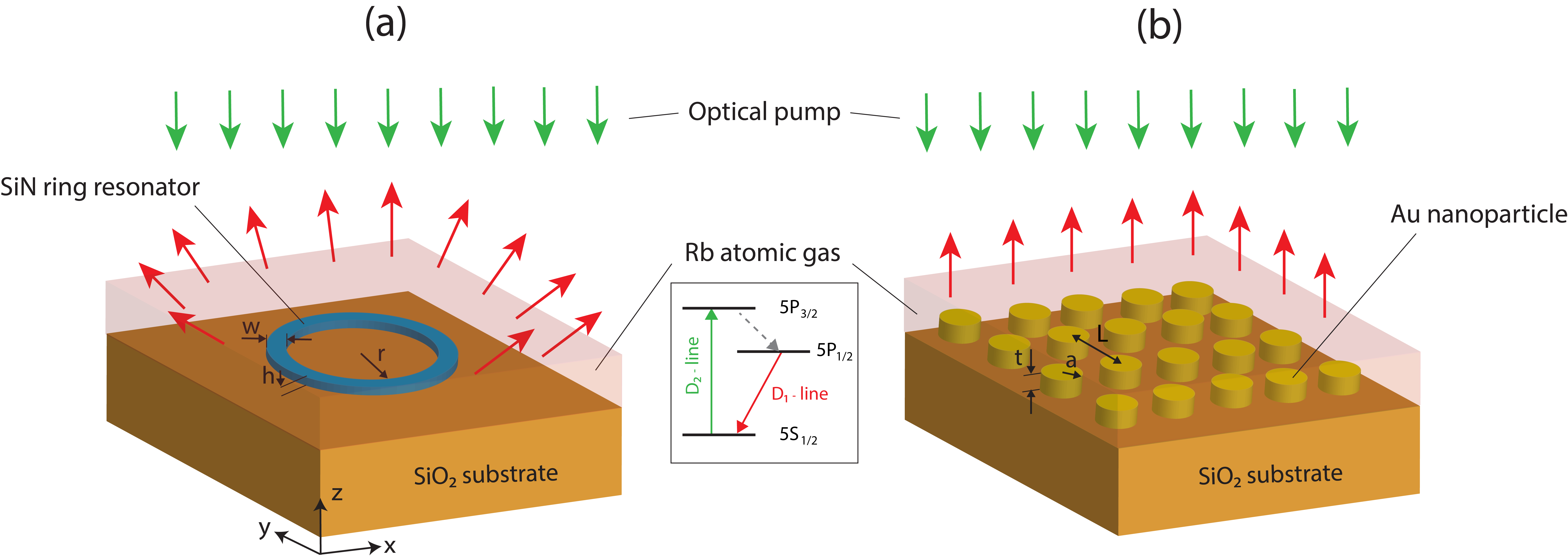}
\caption{\label{Fig1} Schematics of the (a) ring resonator and (b) plasmonic hybrid laser systems studied in this work. The geometrical parameters defining each structure have been indicted in the figures. Top green arrows in both panels represent the external pumping, while the red arrows represent the radiation from each structure. The central inset shows Jablonski diagram of Rb atoms used as active gain medium here.}
\end{figure*}

To model the interaction of this atomic cloud with photonic modes, we adapt the semi-classical approach, where the electromagnetic fields are treated classically and the atoms are considered quantum mechanically. The density matrix is the best way to take into account atomic transitions and decoherence phenomena such as Doppler and transient broadening. Moreover, if the additional decoherence mechanisms such as collision of Rb atoms with buffer gas molecules are considered in Lindblad form, the Liouville equation would properly describe the time evolution of the atomic states when coupled to an arbitrary photon field. In what follows we describe the method for the lasing problem where the gain medium (i.e. Rb vapor and buffer gas mixture) is tread as a 3-level system as shown in Fig.~\ref{Fig1}. Although the final equations and results have been derived for a 3-level gain medium and single mode lasers, the procedure can be generalized for more atomic levels and multi-mode photonic systems. 

The time evolution of the Rb atom density matrix ($\hat{\rho}$) is described via the following Liouville equation:

\begin{eqnarray}\label{Liouville eq}
i\hbar\dot{\hat{\rho}} = [\hat{H}_{A} + \hat{H}_I,\hat{\rho}] + \mathcal{L}[\hat{\rho}]
\end{eqnarray}

Where $\hat{H}_{atom}$ is the free-atom and $\hat{H}_I$ is interaction Hamiltonian describing the coupling between atoms and the cavity field. 

The cavity field $\vec{E}(\vec{r},t)$ on the other hand is determined via Maxwell's equation as:

\begin{eqnarray}\label{Maxwell's eq.}
(\nabla^2-\frac{\epsilon_r(r)}{c^2}\frac{\partial^2}{\partial t^2})\vec{E}(r,t)=\mu_0\frac{\partial^2\vec{P}(r,t)}{\partial t^2}
\end{eqnarray}

In the above equation $\epsilon_r(r)$ is the linear permittivity and in general is a function of position due to the spatial distribution of the refractive index in the photonic structure. $\vec{P}(\vec{r},t)$ is the polarizability of the atomic cloud and for a uniform atomic density of $N$ is related to the induced atom dipole moment as:

\begin{eqnarray}\label{induced dipole moment}
\vec{P}(r,t) =N \braket{\vec{p}(r,t)}=Tr(\hat{p}\hat{\rho}(r,t))N=N(\rho_{21}(r,t)\vec{M} + c.c.)
\end{eqnarray}

where $\vec{M}$ is the transition e-dipole moment between the 1$^{st}$ and 2$^{nd}$ state. 

Here, we implicitly assume that the coherent atom-field interaction only occurs between $\ket{1},\ket{2}$ with the effective transition electric dipole $\vec{M}$. The interaction between $\ket{1},\ket{3}$ is assumed to be incoherent which can be described as an effective pumping rate $R_p$.

Equation~\ref{cavity field2} gives the modes of the cold cavity at $\omega_c$ with the decay rate $\gamma_c$:

\begin{eqnarray}\label{cavity field2}
\vec{E}(r,t)= \vec{E}^+(r,t) + c.c. = \vec{E}_0^+(r)e^{-\gamma_ct}e^{-i\omega_ct} +c.c. 
\end{eqnarray}

\begin{figure*}[t]
\centering
\includegraphics[scale=1]{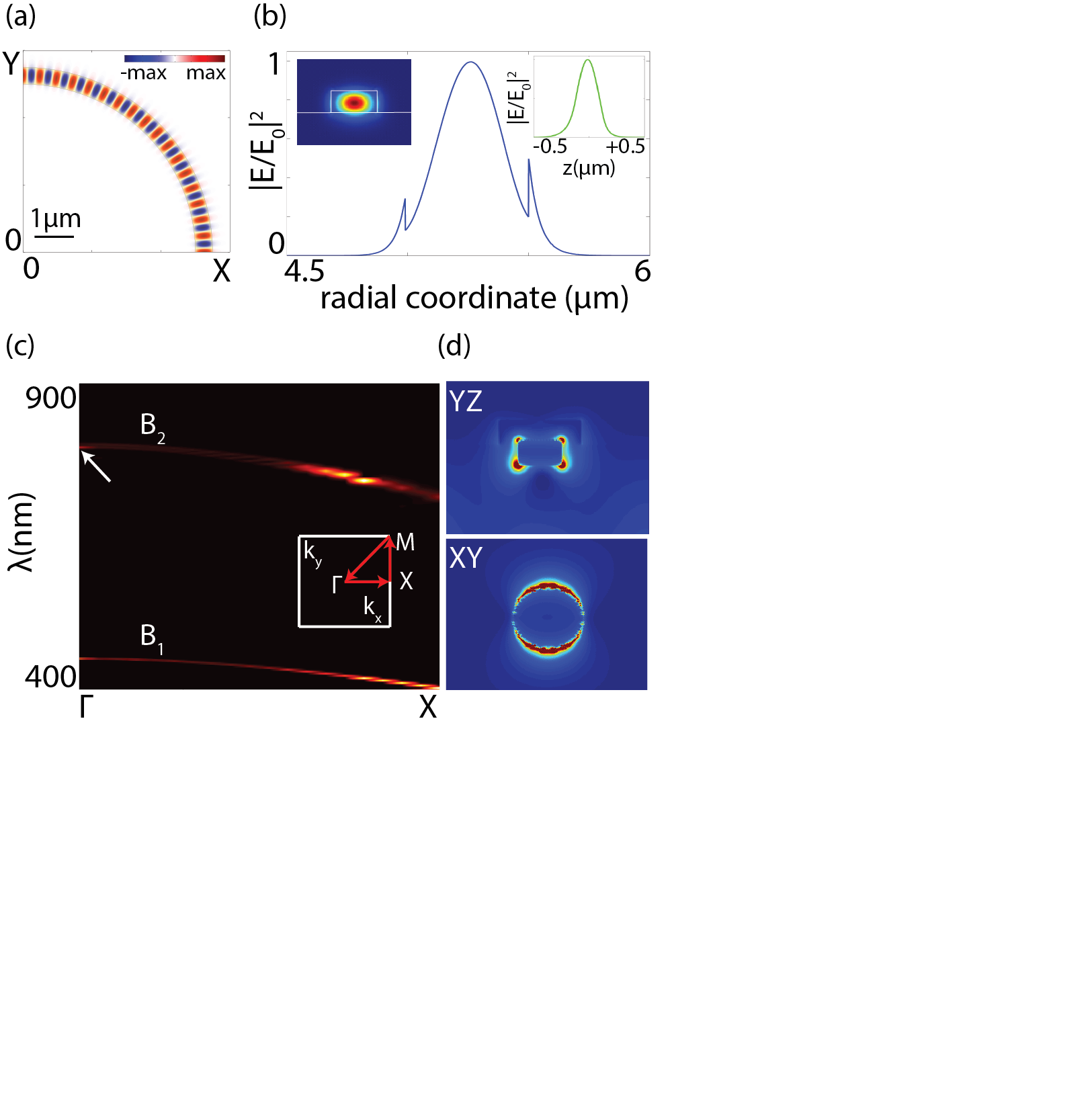}
\caption{\label{Fig2} (a) $xy$ cross-section of the magnetic field distribution ($H_z$) corresponding to the TE-polarized eigenmode supported by the system shown in Fig. 1(a) at $\lambda$=795 nm. The values of the geometrical parameters defining the structure are $r$=5 $\mu$m, $w$=500 nm and  $h$=250 nm. (b) Radial variation of the normalized electric-intensity of the whispering gallery mode supporting the resonance at Rb $D_1$-line. Left inset displays the $yz$ cross-section of the corresponding $E$-field intensity and right inset shows the variation of the $E$-field intensity along $z$-axis normal to the ring plane. (c) Band diagram of the plasmonic lattice along $\Gamma X$ direction in the first Brillouin zone showing two band ($B_1 , B_2$) within the visible and infrared range. (d) The $yz$- (top) and $xy$-cross section (bottom) electric field intensity of the mode resonating at about 795 nm indicated by a white arrow in panel (c).}
\end{figure*}

After properly incorporating the finite-lifetime of each state and in the validity realm of rotating wave approximation (RWA), one would have the following sets of equations of motion for the density matrix elements:

\begin{subequations}\label{density matrix evolution}
\begin{align}
\dot{\rho_{33}}(r,t) &=-(\gamma_{31}+\gamma_{32})\rho_{33}(r,t)+R_p(\rho_{11}(r,t)-\rho_{33}(r,t))\\
\dot{\rho_{22}}(r,t) &=\gamma_{32}\rho_{33}(r,t)-\gamma_{21}\rho_{22}(r,t) \nonumber \\
&+\frac{1}{i\hbar}(\vec{M}\cdot\vec{E}^-(r,t)e^{+i\omega_c t}\rho_{21}(r,t)-c.c.)\\
\dot{\rho_{11}}(r,t) &= \gamma_{31}\rho_{33}(r,t)+\gamma_{21}\rho_{22}(r,t)-R_p(\rho_{11}(r,t)-\rho_{33}(r,t)) \nonumber \\
&-\frac{1}{i\hbar}(\vec{M}\cdot\vec{E}^-(r,t)e^{+i\omega_c t}\rho_{21}(r,t)-c.c.)\\
\dot{\rho_{21}}(r,t) &=-(i\omega_a+\gamma_\vert)\rho_{21}(r,t)+\frac{1}{i\hbar}\vec{M}^*\cdot\vec{E}^+(r,t)e^{-i\omega_c t}w
\end{align}
\end{subequations}

where $\gamma_\vert = \gamma_{32} + \frac{\gamma_{21}}{2}$ is the transverse decoherence rate taking into account all the de-coherence effects, including the collision with the buffer gas.

In contrary to many of the large cavities where the uniform field approximation can be used, in nano-photonic and plasmonic systems one cannot ignore the spatial variation of the electromagnetic fields inside the cavity as the field features vary below or on the order of the wavelength. That makes the density matrix elements functions of both time and position. In other words the above sets of equations have to be solved simultaneously at each point of the space to properly capture all the atomic and photonic features. To overcome this, we project the fields on the basis of the cold cavity modes. 

Therefore the modified cavity field can be rewritten as:

\begin{eqnarray}\label{modified cavity field}
\vec{E}(r,t)={E}_0(\vec{f}_0^+(r)a^+(t)e^{-i\omega_ct}+c.c.)
\end{eqnarray}

where we have defined $E_0 = max[\vec{E}_0^+(r)|]$ and the dimensionless vectors of $\vec{f}_0^\pm(r)=\vec{E}_0^\pm(r)/E_0$.

For the observables appearing as diagonal elements of the density matrix the spatially averaged parameters would be determined as:
\begin{eqnarray}\label{average density}
\braket{\rho_{ii}(t)}=\frac{\int_{atom} dv \rho_{ii}(r,t)|\vec{f}_0^+(r)|^2}{\int_{atom} dv |\vec{f}_0^+(r)|^2}
\end{eqnarray}

The subscript \emph{atom} in the above integral indicates an integration over the active region where the polariazable atoms are present. For the off-diagonal elements related to the coherent interactions we employ the following approximation within the active region:

\begin{eqnarray}\label{off-diagonal app}
\vec{M}\rho_{21}=|\vec{M}|\vec{f}_0^+(r)e^{-i\omega_at}\tilde{\rho}_{21}(t)
\end{eqnarray}
 
Using the above definitions the equations of motion in eq.~\ref{density matrix evolution} combined with eq.~\ref{Maxwell's eq.} for the cavity field could be rewritten as:

\begin{subequations}\label{CM equations}
\begin{eqnarray}
\frac{d\braket{\rho_{33}}}{dt}=-(\gamma_{31}+\gamma_{32})\braket{\rho_{33}} + R_p(\braket{\rho_{11}}-\braket{\rho_{33}})
\end{eqnarray}
\begin{eqnarray}
\frac{d\braket{\rho_{22}}}{dt}=\gamma_{32}\braket{\rho_{33}} -\gamma_{21}\braket{\rho_{22}} +\frac{\zeta}{i\hbar}(\tilde{\rho}_{21}a^-(t)e^{i\delta t}-c.c.)
\end{eqnarray}
%
%
\begin{eqnarray}
\frac{d\tilde{\rho}_{21}}{dt}=-\gamma_\vert\tilde{\rho}_{21}+\frac{1}{i3\hbar}|\vec{M}|E_0\braket{w} a^+(t)e^{-i\delta t}
\end{eqnarray}
\begin{eqnarray}
\frac{d}{dt}a^+(t)=-\gamma_c a^+(t)+[i\frac{\omega_a}{2}\tilde{\rho}_{21}-\frac{d}{dt}\tilde{\rho}_{21}]\mu_0 Nc^2\xi\frac{\omega_a}{\omega_c} e^{i\delta t}
\end{eqnarray}
\end{subequations}

where $\delta = \omega_c-\omega_a$ is the detuning between the cavity resonance and the atomic frequency and $\braket{w}=\rho_{22} - \rho_{11}$ is the population difference. 
Moreover, $\zeta , \xi$  describe the effective interaction of the field with active region and are defined as follows:

\begin{subequations}\label{zeta1,2}
\begin{align}
\xi =\frac{|\vec{M}|}{E_0} \frac{\int_{atom} dv |\vec{f}_0^+(r)|^2}{\int_{space} dv \epsilon(r)|\vec{f}_0^+(r)|^2}=  \frac{|\vec{M}|}{E_0}\zeta_1
\\
\zeta =|\vec{M}|E_0\frac{\int_{atom} dv |\vec{f}_0^+(r)|^4}{\int_{atom} dv |\vec{f}_0^+(r)|^2}=|\vec{M}|E_0\zeta_2
\end{align}
\end{subequations}

The above equations are very general and can be used to describe the interaction of the atom cloud with arbitrary fields of a nano-device in any configuration.
\begin{figure*}[t]
\centering
\includegraphics[width=\linewidth]{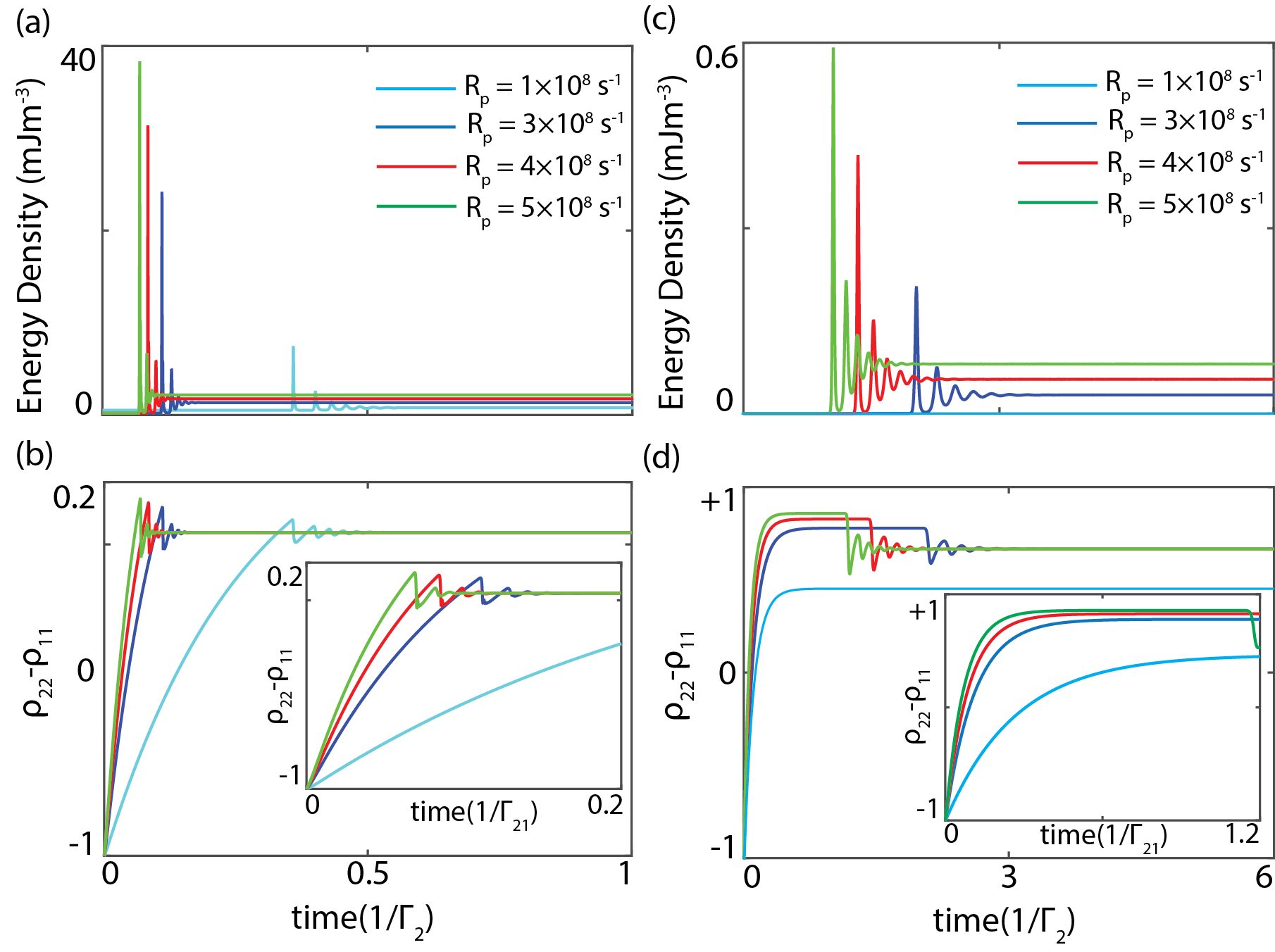}
\caption{\label{Fig3} (a) Variation of the electric field energy density and (b) population difference between $\ket{2}$ and $\ket{1}$ as a function of time in a ring resonator at various pumping rates. 
Temporal evolution of (c) the electric field energy density and (d) population difference between $\ket{2}$ and $\ket{1}$ for the plasmonic lattice at few different pumping rates. The inset figures in panel (b) and (d) show the dynamical behavior of the population difference at early times, illustrating the fast increase of the excited state population. For both resonators the Rb gas density is assumed to be $N = 10^{22}~N^{-3}$. The normalizing factor $\Gamma_{2}$ is the natural decay rate of $2^{nd}$-level as 36$\times 10^{6} ~ s^{-1}$. }
\end{figure*}
To solve for the lasing features as we are interested in this paper, eq.~\ref{CM equations} must be solved with the following initial conditions when a proper pumping exists.
~\footnote{To solve the equations numerically one has to seed the $a^{+}(0)$ at a very small value. This insures that the system of equations will evolve in time and would not stuck at initial conditions mentioned in eq.~\ref{initial conditions}. The final results are checked to be completely independent from this seeding value.}

\begin{eqnarray}\label{initial conditions}
\rho_{11}(0)=1,  \rho_{22}(0)=\rho_{33}(0)=a^+(0)=0, R_p\neq 0
\end{eqnarray}

Finally, note that all of the unknown variables in eq.~\ref{CM equations} are time-dependent only, making the atom-cavity interaction problem more tractable numerically while providing more physical insight.

\section{Results and discussion}

\subsection{Description of the analyzed systems and passive response}

Figure~\ref{Fig1} shows schematics of the two hybrid integrated laser systems under study. The first configuration (Fig.~\ref{Fig1}(a)) consists of a dielectric SiN ring of inner radius $r$, height $h$ and width $w$, lying on top of a SiO$_2$ substrate. The dielectric ring is embedded in a thermal cloud of Rb atoms with the density of $N$, and ethane molecules . The Rb atoms are modeled as three-level systems, excited with an incoherent pump at $\lambda_{D_2}$=780 nm. The fast and nonradiative transition between 5$P_{3/2}$ and 5$P_{1/2}$ states (mainly achieved via collision with buffer gas molecules such as He or ethane) enables the population inversion between this level and the ground state to be built up at $D_1$-transition with $\lambda_{D_1}$=795 nm (see Jablonski diagram of the inset in Fig.~\ref{Fig1}). 

The second structure is a periodic lattice of Au (gold) cylindrical nanoparticles (Fig.~\ref{Fig1}(b)). The lattice constant is given by $\Lambda$, and the radius and height of each of the nanoparticles is $r$ and $t$, respectively. The rest of the details are the same as those described above.

To get physical insight into the above described route to lasing action, we focus first on analyzing the cold-cavity features in the absence of the Rb atoms. Figures~\ref{Fig2}(a) and (b) summarize the results of the ring resonator. In these calculations we have assumed refractive indexes of $n=2.00$ and $n=1.45$ for the SiN and SiO$_2$ regions of the structure, respectively and with the geometrical parameters of,  $r$=5$~\mu$m, $w$=500 nm, and $h$=250 nm. These values have been obtained by optimizing simultaneously the values of the ring radius and the azimuthal order of the whispering gallery modes (WGMs), until the mode resonances matches to the $D_1$-line of Rb, i.e. $\delta$=0. (For further details on the optimization process refer to the Supplementary Information, where the effects of $r$ and $m$ on the modal resonance have been investigated further). The effect of a non-zero detuning of the WGM resonance with respect to lasing transition will be discussed later in the paper.

Figure~\ref{Fig2}(a) shows $H_z$ profile of a TE-polarized eigenmode supported by the optimal structure at $\lambda=795$ nm and $m=68$. The corresponding electric-field intensity distribution is shown in Fig. ~\ref{Fig2}(b). As can be seen, most of $E$-field is concentrated inside the dielectric ring, but still there are fractions of the field extending beyond the ring interacting with the atomic cloud.

Similarly, Fig.~\ref{Fig2}(c) and (d) depicts the results of the plasmonic lattice in Fig.~\ref{Fig1}(b), when $\Lambda=450$ nm, $t=60$ nm and $a=45$ nm. The substrate is assumed to be SiO$_2$ and the metallic nano-particles are described with Johnson and Christy empirical data.
Fig.~\ref{Fig2}(c) displays the band diagram of the periodic lattice along the $\Gamma-X$  direction of the First Brillouin Zone (see the inset of Fig. ~\ref{Fig2}(c)). This band diagram was obtained through a Fourier transform analysis of the slowly-decaying eigenmodes of the system after being excited by a set of point dipoles located at random positions (we carried out this analysis using the commercial implementation of the finite-different-time-domain method provided by Lumerical, see Supplementary Information for details). 

Two plasmonic bands exist in the spectrum. The mode of our interest is the one at $\lambda=$795.18 nm for $k_x=0$ (see white arrow in main panel of Fig.~\ref{Fig2} (c)). As for the ring, the geometrical parameters of the plasmonic lattice have been optimized so that the \emph{lattice plasmon resonance} (formed by the hybridization of the diffracted lattice modes with the localized surface plasmon resonances of each metallic nano-particle) appears as close as possible to $D_1$ transition of Rb. Due to their larger $Q$-factor, in comparison with other plasmonic resonances, lattice plasmons have been successfully used for realizing plasmonic nano-lasers in the past~\cite{Zhou2013,Vanbeijnum2013,Hakala2017}. Figure~\ref{Fig3}(d) shows the cross-sections of the $E$-field distribution corresponding to this lattice mode illustrating large-field enhancements associated to this resonance. Finally, we highlight that, in contrast to the ring resonator, we have not been able to fully tune the lattice plasmons to atomic resonance while keeping the geometrical parameters in a reasonable range. Next section discusses the implications of this on the laser performance and dynamics.

\subsection{Lasing dynamics and steady-state laser characteristics}
In this section we consider the behavior of the resonators designed in the previous section in the vicinity of the thermal cloud and buffer gas mixture.  The buffer gas leads to a rapid dephasing of both Rb isotopes, producing a homogeneous
broadening on the order of several tens of GHz. This strongly depends on the buffer gas pressure. Here we considered a 12 GHz line width due to this phenomenon. We further assume that these buffer gas-induced dephasing rates are the same for all transitions~\cite{Yablon16}. Moreover, $38\times 10^{6}~s^{-1}$ and $36\times 10^{6}~ s^{-1}$ are the natural decay rates of $\ket{3}$ and $\ket{2}$, respectively. Finally, the transition dipole moment of Rb is $|\vec{M}| = 25\times 10^{-30}$ Cm in the following calculations~\cite{Steck01}. 

Figure~\ref{Fig3} summarizes the dynamical characteristics of the coupled atom-cavity system. Figure~\ref{Fig3}(a) and (c) show the time evolution of the spatially averaged electric energy density of the ring resonator and plasmonic lattice, respectively.
The results has been shown for four different pumping rates.
In both systems, we obtain the canonical features of lasing dynamics, consisting of a series of sudden spikes in the field energy signal that settle down to a non-zero value in steady-state, indicating that both systems can indeed lase at large enough pumping rates. This is the first main results of this work demonstrating the possibility of obtaining coherent radiations in such hybrid systems. In particular, as seen in Figs.~\ref{Fig3}(a) and (c), the micro-ring cavity displays shorter lasing onset times and faster oscillations than the plasmonic lattice. At $R_p = 5\times 10^{8}~s^{-1}$, the ring resonator takes 3.3 $ns$ to reach to the steady state, while this time increases to 65 $ns$ in the plasmonic lattice which is about 20 times longer.

\begin{figure}[t]
\centering
\includegraphics[scale=1]{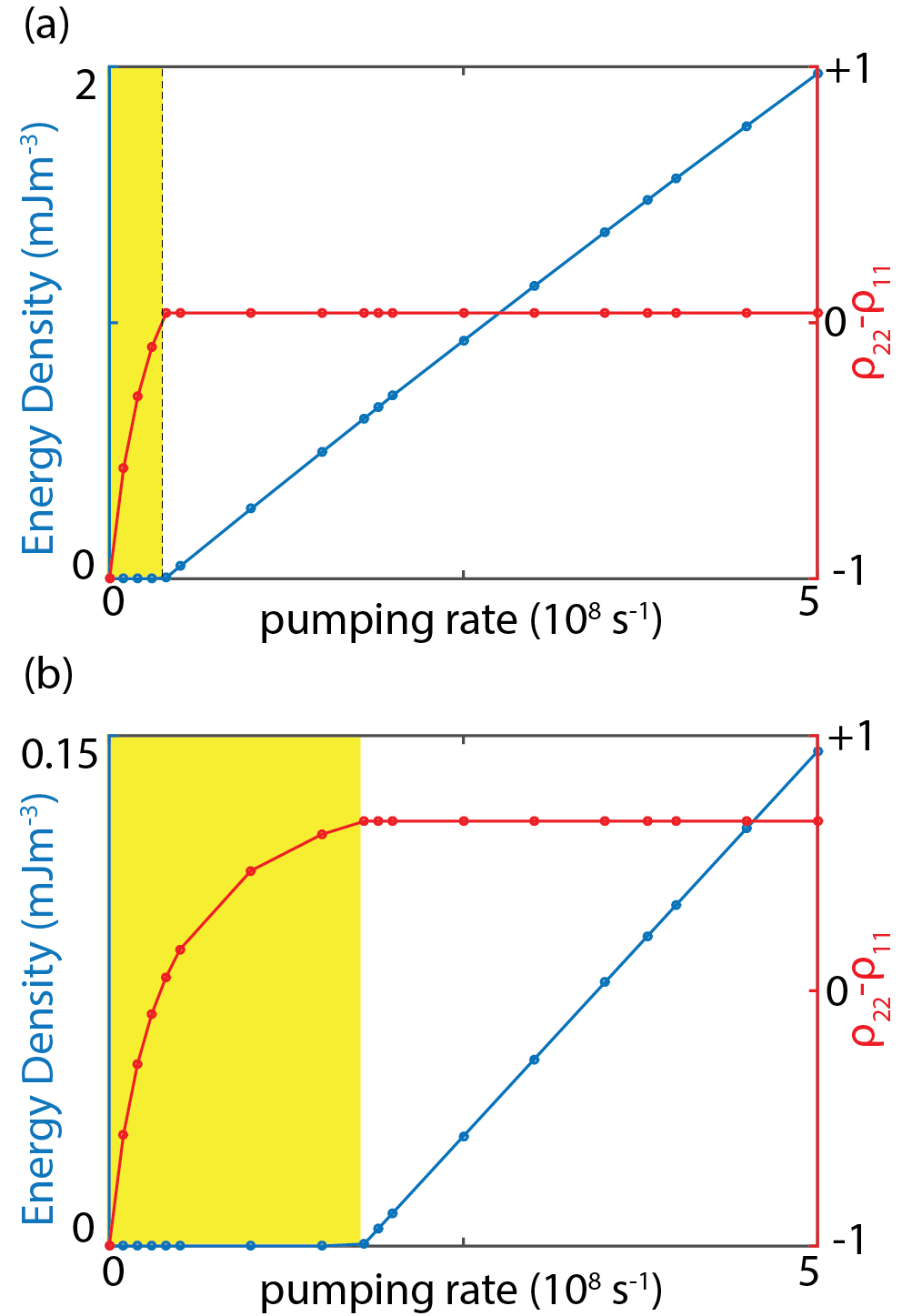}
\caption{\label{Fig4} Dependency of the laser steady-state features, i.e. energy density (blue line) and the population difference (red line), as a function of pumping rate ($R_p$) for the (a) ring resonator and (b) plasmonic lattice. The yellow region in both panels highlights the below-threshold region of the nano-lasers where lasing does not happen.}
\end{figure}

\begin{figure*}[t]
\centering
\includegraphics[width=\linewidth]{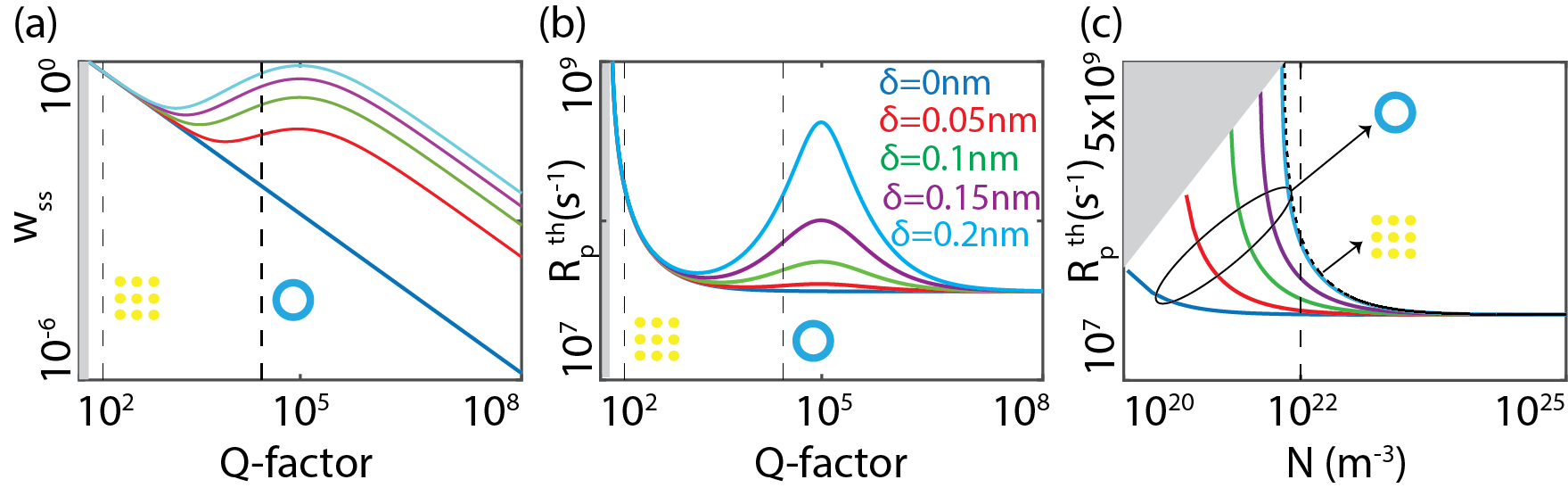}
\caption{\label{Fig5} (a) Variation of the required population inversion leading to a coherent lasing emission in a nano-laser as a function of its quality factor ($Q$) when the density of the atoms in the active region in assumed to be fixed at $N=10^{22}~m^{-3}$ and active medium fractional ratio is 4$\%$. Each line corresponds to different detuning $\delta$ between the cavity and the atomic transition resonance. The gray area in the low-Q part indicates the non-lasing regime and the vertical dashed lines show the behavior of $w_{ss}$ for the plasmonic lattice and micro-ring studied in this work.
(b) The effect of cavity quality factor on the pumping threshold of the nano-laser for five different atom-cavity detunig. The black dashed lines show the dependency of the plasmonic lattice and ring resonators studied in this work. (c) The influence of atom density in controlling the laser threshold in a ring resonator laser with $Q=3\times 10^4$ as designed and investigated in this work for various atom-cavity detuning. The dashed line shows the tendency for the plasmonic lattice. The vertical dashed line shows the atomic density considered in in this paper.}
\end{figure*}

To investigate the link between the field oscillations and the population of the electronic levels of Rb atoms, Figs.~\ref{Fig3}(c) and (d) show the corresponding time evolution of the spatially-averaged population inversion ($\rho_{22} - \rho_{11}$). In both cases, before the first lasing spike occurs, the averaged population inversion rapidly grows with time (see insets of Figs.~\ref{Fig3}(b) and (d)). This corresponds to the regime where the population of $\ket{2}$ is increasing (the system is \emph{accumulating} population inversion). When the population difference between $\ket{2},\ket{1}$ becomes large enough so its associated optical gain can overcome all the de-coherence phenomena, the first burst of laser radiation takes place, as can be observed in Figs.~\ref{Fig3}(b) and (d) for larger values of $R_p$. This burst leads to a significant depletion of the population inversion (a significant amount of the upper-level population of the laser transition decays via stimulated emission), leading to a dramatic drop of the laser signal. After that, it starts a subsequent recovery of the populations inversion, until enough population inversion is accumulated again and a second spike of the lasers signal occurs. This is  accompanied by the corresponding drop in the population inversion. This series of bursts and subsequent drops of the populations inversion takes place sequentially (for larger times smoother spikes and drops of the lasing signal and population inversion are obtained) until the steady-state is reached. Note that the steady-state value of the population inversion required for lasing depends on the de-coherence phenomena in the gas as well as the optical cavity. Therefore, After reaching the threshold the optical gain is clamped and the population inversion remains fixed.

The better the cavity the smaller the de-coherence, hence smaller gain and population inversion are required. For the ring resonator with quality factor $Q\approx 3\times 10^4$ the required population inversion of lasing is $\approx 0.04$. While due to the lower quality factor of the lattice plasmon resonance $\approx 211$ this value increases to 0.66.

Fig.~\ref{Fig4} (a) and (b) show the steady-state behavior of electric-file energy density and the population difference as a function of $R_p$ for the micro-ring and the plasmonic lattice, respectively. The observed linear dependence above threshold confirms that the two considered configurations are indeed lasing. In addition, as can be seen the plasmonic lattice starts to lase at the pumping threshold of $R_p^{th} = 1.9\times 10^{8}~s^{-1}$, while the same parameter reduces to $R_p^{th}\approx 4.1\times 10^7 s^{-1}$ for micro-ring, which is about 5 times smaller.

\subsection{Simplified model}
To develop further insight to the performance of these hybrid lasers, we use a simplified analytic model that captures the main lasing characteristics of these systems. From eq.~\ref{CM equations} it is straightforward to calculate the required population difference between $1^{st}$ and $2^{nd}$-levels to overcome all decoherence phenomena in the atom-cavity system and achieve a sustainable lasing oscillation as:

\begin{equation}\label{steady-state population inversion}
w_{ss} = \frac{6\hbar[(\gamma_c\gamma_\vert + \delta_1\delta_2)Re(\zeta_1)+(\delta_1\gamma_\vert - \delta_2\gamma_c)Im(\zeta_1)]}{(\omega_a + 2\delta_2)N|\vec{M}|^2\mu_0c^2\omega_a/\omega_c|\zeta_1|^2}
\end{equation}

This population inversion can be achieved with an incoherent pump at the rate of ~$R_p^{th} = \nobreak \gamma_{21}(1+\nobreak w_{ss})/(1-w_{ss})$.

In eq.~\ref{steady-state population inversion}, $\delta_{1,2}$ represent the frequency pulling effect due to a non-zero frequency detuning between cavity and atom resonances, defined by $a^+(t) \rightarrow a^+(t)e^{+i\delta_1 t}$ and $\tilde{\rho}_{21}(t) \rightarrow \tilde{\rho}_{21}(t)e^{-i\delta_2 t}$. As expected $w_{ss}$ is dependent on the longitudinal and transverse decay rates ($\gamma_c,\gamma_\vert$), frequency detuning ($\delta_{1,2}$), atom-field coherent coupling strength set by $M$, atom density ($N$), and the ratio of the active region volume to the cavity mode volume $\zeta_1$. In a nano-photonic cavity $\zeta_1$ and $\delta_{1,2}$ are strongly dependent on the device features, hence $w_{ss}$. However, a good approximation to the numerical results can be obtained by assuming a reasonable range for $\zeta_1$. Based on various studies on these two classes of cavities we found that $\zeta_1 = 0.04$ is a proper approximate value for micro-ring as well as the plasmonic lattice. (See Table 1 of Supplementary Information) 

For the fixed atomic density of $N=10^{22}~m^{-3}$, Fig.~\ref{Fig5}(a) and (b) show the behavior of required population inversion and corresponding pumping rate as a function of cavity quality factor, respectively. For better understanding, the behavior has been investigated for different values of the detuning between the cavity resonance and the atomic transition, ranging from $\delta$=0.0 to $\delta$=0.20 nm. The two vertical dashed lines show the trend for plasmonic lattice and ring-resonator with $Q_P \approx 211$ and $Q\approx 3\times 10^4$, respectively. 

As seen in Fig.~\ref{Fig5}(a) in a bad cavity regime ($Q\le 135$), the required population inversion is $\ge 1$ , implying that the achievable gain in such a system is not large enough to overcome all the decoherence, hence no laser radiation can be observed within this range. The corresponding pumping rate of this region is also very high as can be seen in Fig.~\ref{Fig5}(b). This lasing-forbidden region is highlighted in gray in Fig.~\ref{Fig5}(a) and (b).

For larger moderate cavities ($Q\le 500$) there is a intermediate region, where the inversion and pumping thresholds are almost insensitive to the atom-cavity detuning. That is due to the fact that $\gamma_c$ is still too large compared to the transverse decay that the detuning is mainly leads to a cavity pulling and not an atom pulling (i.e. $\delta_0=0$), making $w_{ss}$ independent of $\delta$. As shown in Fig.~\ref{Fig5}(a) and (b), plasmonic lattice falls within this range, hence any further optimization of its geometrical features to decrease the detuning does not have a substantial effect in reducing the required gain or pumping threshold. 

For larger quality factors the detuning always increases the required inversion and pumping rate as can be seen in Fig.~\ref{Fig5}(a) and (b), respectively. The micro-ring is a cavity within this range where one can see the substantial sensitivity in its steady-state features as a function of detuning.

Unlike the monotonic dependency to the detuning, the dependency to the cavity quality-factor is only monotonic for $\delta=0$, where there is no pulling effect. For other cases the required gain at first increases with increasing the quality factor and after reaching to a maximum, monotonically drops for better cavities. This behavior is a direct consequence of frequency pulling effect. While the bad cavity tends to only detune the cavity mode resonance (i.e. $\delta_2 = 0$), a good cavity would alter the atomic resonance ($\delta_1= 0$). For regions in between the both of the atom and cavity resonance would be modified depending on the relation between $\gamma_c,\gamma_\vert$ leading to a non-monotonic dependency to $Q$ as in Fig.~\ref{Fig5}(a) and (b).

Finally, Fig.~\ref{Fig5}(c) illustrates the monotonic decrease of the required pumping rate with atom density $N$ when $Q=3\times 10^4$ as for the micro-ring. The vertical dashed line shows the density used throughout this paper.
Different line correspond to various detuning. According to the previous discussion, for a fixed atom density the required inversion of each atom, hence the pumping rate, should be larger to overcome the detuning loss effect. Moreover, if the detuning is fixed one can increase the cloud density to need less stimulated emission rate from each individual atom, as suggested by the monotonically decreasing trend in Fig.~\ref{Fig5}(c) for each fixed detuning. The gray shaded area on the top-left corner indicates the parameter ranges where no lasing can be achieved either due to the small atom density, or the large detuning effect. The overlaid black dotted line shows the behavior of the plasmonic lattice investigated in this paper. As can be seen in spite of very different nature lattice resonance modes compared to the ring resonator case they both follow the same tendency. This might seem a surprising result as the quality factor of the plasmonic cavity is orders of magnitude smaller than the ring resonator.  However, it must be noted that the lattice mode volume is much smaller than the micro-ring's, leading to comparable values for cavity mode, atom-radiation fractional overlap.

\section{Conclusions}
In this paper we report a novel route for realizing a coherent light source using hybrid systems based on incorporating an alkali-buffer gas mixture in integrated nanophotonic structures. To study this class of systems, we have developed a powerful and efficient theoretical method based on the density-matrix formalism, which has allowed us to characterize both the dynamic and steady-state features of the nonlinear light-matter interaction occurring in the considered structures. Using this formalism we have systematically studied two realistic configurations based on an optically-pumped Rb-ethane gas mixture into two types of numerically optimized nanophotonic systems, i.e., a dielectric ring resonator and plasmonic lattice. For both classes systems, we have shown how by tuning one of their corresponding photonic resonances to a suitable transition of the atomic cloud, it is possible to enable emission of coherent radiation at low power levels. In addition, using a simplified model, we have provided general guidelines for the design of the studied hybrid microlasers.

To the best of our knowledge, this is the first study of lasing action in such active hybrid systems. Atomic-photonic hybrid systems are rather novel and unexplored platform, and we believe the present study could stimulate further research aimed an exploiting their unique quantum-optical properties. Thus, for instance, the study of the response of these hybrid systems in different light-matter coupling regimes is a promising future research direction that will provide essential information for realizing other quantum optical  phenomena in this platform.

\section*{Acknowledgments}
H. A. acknowledges the insightful and stimulating discussions with Teri Odom and Selim Shahriar, and the financial support from MRSEC under grant DMR-1121262.

J.B.-A. acknowledges financial support by Spanish MINECO under grant MAT2015-66128-R (MINECO/FEDER).

\bibliography{ref}

\end{document}